\documentclass[12pt]{iopart}
\usepackage{graphicx}
\newcommand{\sinc}{{\rm sinc}}
 
\begin{document}

\title[Thomson backscattering from laser-generated, relativistically moving electron layers]{Thomson backscattering from laser-generated, relativistically moving high-density electron layers}

\author{A~Paz$^{1,2}$, S~Kuschel$^1$, C~R\"odel$^{1,2}$, M~Schnell$^{1,2}$, O~J\"ackel$^{1,2}$, M~C~Kaluza$^{1,2}$ and G~G~Paulus$^{1,2}$}

\address{$^1$Institut f\"ur Optik und Quantenelektronik, Friedrich-Schiller-Universit\"at Jena, Max-Wien-Platz 1, 07743 Jena, Germany}
\address{$^2$Helmholtz Institut Jena, Fr\"obelstieg 3, 07743 Jena, Germany}
\ead{athena-evalour.paz@uni-jena.de}

\begin{abstract}
We show experimentally that XUV radiation is produced when a laser pulse is Thomson backscattered from sheets of relativistic electrons which are formed at the rear-surface of a foil irradiated on its front side by a high-intensity laser.  An all-optical setup is realized using the Jena Titanium:Sapphire TW laser system (JETI). The main pulse is split into two pulses: one to accelerate electrons from thin aluminium foil targets to energies of the order of some MeV and the other, counterpropagating probe pulse is Thomson-backscattered off these electrons when they exit the target rear side. The process produced photons within a wide spectral range of some tens of eV as a result of the broad electron energy distribution. The highest scattering intensity is observed when the probe pulse arrives at the target rear surface 100 fs after the irradiation of the target front side by the pump pulse, corresponding to the maximum flux of hot electrons at the interaction region. These results can provide time-resolved information about the evolution of the rear-surface electron sheath and hence about the dynamics of the electric fields responsible for the acceleration of ions from the rear surface of thin, laser-irradiated foils.
\end{abstract}

\submitto{New Journal of Physics}
\maketitle

\section{Introduction}
\label{sec:Introduction}
Thomson scattering of intense laser pulses from relativistic electrons has the potential to produce high-quality, ultrashort pulses of well-collimated x-ray radiation \cite{Esarey,Lau,Lan}. Such an x-ray source would have a great impact on biomedical imaging, spectroscopy and the analysis of ultrafast structural dynamics \cite{Brown,Anderson}.  When implemented using an all-optical setup such that the laser pulse scatters from laser-accelerated electrons, it possesses the additional advantage of having a lower electron energy requirement to generate a given photon energy compared to undulator-based radiation sources at conventional electron accelerators such as synchrotrons. This is due to the fact that the laser wavelength is orders of magnitude shorter than the undulator magnet period \cite{Esarey}.  It is also compact in size and provides better control over the synchronization as compared to setups where a laser pulse is scattered from electrons accelerated e.g. by a conventional linear accelerator \cite{Karagodsky}. Furthermore, it can be used as an in-situ time-resolved diagnostic for the laser-matter interaction \cite{Liesfeld}.  However, despite these benefits, only a few experimental investigations have been conducted on the subject using a purely laser-based setup \cite{Schwoerer}.  

In this paper we describe an all-optical setup realizing Thomson backscattering using a 2.4 $\mu$m thick aluminium foil target. In the course of the discussion we first give a summary of the predicted properties of the backscattered radiation relevant to this experiment. We then provide a brief description of the electron acceleration from a thin foil target and the evolution of the hot-electron sheath. This is followed by the discussion of the optimization of the spatio-temporal overlap of the counterpropagating pulses at the interaction region. Finally, we present the spectrum of the generated XUV radiation.  

\subsection{Properties of the Thomson backscattered photons} 
   
\begin{figure}
\begin{center}  
	\includegraphics[width=0.6\textwidth]{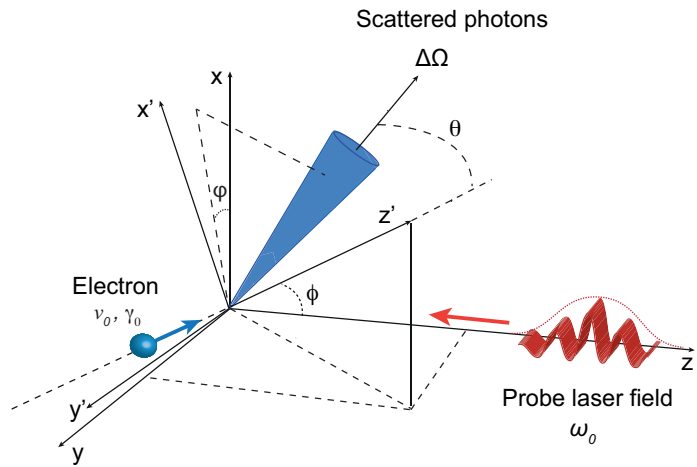}
\end{center}
	\caption{\label{thomson}The interaction geometry of Thomson backscattering. An electron propagating along z'  with a Lorentz factor $\gamma_{0}$ collides with a laser pulse propagating in -z-direction emitting radiation at Doppler upconverted frequencies. During the interaction, the z' and z axes enclose the angle $\phi$ \cite{Ride}.} 
\end{figure}

The conversion of visible laser light into x-ray radiation through Thomson backscattering involves the collision of the laser photons with relativistic electrons which may have been generated from an intense laser pulse. In this scenario the probe laser pulse and its wavelength appear length contracted by the initial relativistic factor $\gamma_{0} = (1-\beta^{2}_{0} )^{-1/2}$ of the electron beam in its frame of reference, where $\beta_{0} = v_{0}/c$ is the initial electron velocity (prior to the interaction with the probe pulse) normalized to the speed of light $c$.  Upon interaction, the electrons oscillate in the field of the laser pulse leading to the emission of radiation which is Doppler upshifted when measured in the laboratory frame [see \Fref{thomson}].  Consequently, the peak energy of the backscattered radiation in the laboratory frame dominantly scales as  $\gamma_{0}^2$ and is given by 
\begin{equation}
	\hbar\omega_{\rm x}  = \frac{2\left(1+\cos\phi\right)\gamma^2_{0}}{1+a^2_{0}/2 +\gamma^2_{0}\theta^2} n \hbar\omega_{0}\label{Ex}
\end{equation}
where $\phi$ is the electron-probe beam interaction angle, \emph{n} is the harmonic number of the emitted radiation, $\hbar$ is Planck's constant, $a_{0} = eE_{0}/(m_{\rm e}c\omega_{0}$) is the normalized vector potential of the probe laser field, $e$ and $m_{\rm e}$ are the charge and rest mass of the electron respectively, $E_{0}$ and $\omega_{0}$ denote the probe laser electric field amplitude and frequency, respectively and $\theta$ is the angle between the electron propagation direction and observation axis \cite{Debus,Wu}.  The $1+a^2_{0}/2$ factor accounts for the changes in the longitudinal velocity of the electron as it traverses the probe pulse due to ponderomotive effects - resulting in a shift of the peak to a lower energy in the intense field limit ($a_{0}^2\geq 1$)  \cite{Chang,Krafft}. Moreover, \eref{Ex} implies that the upshift factor varies from $2\gamma_{0}^2$ for transverse interactions ($\phi = 90^{\circ}$) to a maximum of  $4\gamma_{0}^2$ for a head-on scattering geometry ($\phi = 0^{\circ}$).  It is also important to note that in \eref{Ex} it is assumed that the electrons  are highly relativistic with $\beta_{0} \approx 1$ where the radiation is mostly backscattered into small angles $\theta^2 \ll 1$ \cite{Esarey}.  This classical description of Thomson scattering is valid as long as the incident photon energy is less than the electron rest mass in its rest frame \cite{Meyerhofer}.  However, if the energy of the  incident photon  becomes comparable to the electron energy, the electron recoil is no longer negligible and the interaction can be described in the framework of Compton scattering \cite{Bock}. The scattered photon energy from this inelastic process is given by   
\begin{equation}
	\hbar\omega_{\rm x}  = \frac{\left(1+\beta_{0}\cos\phi\right)}{1-\beta_{0}\cos\theta + \left(\hbar\omega_{0}/\gamma_{0}m_{\rm e}c^2\right)\left(1-\cos\Delta\Theta\right) }  \hbar\omega_{0}\label{Compton}
\end{equation}
where $\cos\Delta\Theta = \cos\phi\cos\theta-\sin\phi\sin\theta\cos\varphi$ and $\Delta\Theta$ is the angle between incident and scattered photons \cite{Priebe, Chouffani}. Note that under the Thomson scattering condition $\gamma_{0}\hbar \omega_{0}\ll m_{\rm e}c^2$ \cite{Meyerhofer},  \eref{Compton} reduces to \eref{Ex}.  Furthermore, the classical Thomson formula is recovered by applying the Thomson scattering condition on \eref{Compton} and the limits $\phi = 0^{\circ}$ and $\theta = 0^{\circ}$ .    

The energy radiated by a single electron per frequency interval d$\omega$ and per solid angle d$\Omega$ during the interaction can be computed from the Lienard-Wiechert potentials \cite{Jackson}.  For a head-on scattering geometry of a relativistic electron  with a linearly polarized probe laser pulse of arbitrary intensity, the radiation along the backscattering direction ($\theta = 0^{\circ}$) occurs only at the odd harmonics \cite{Esarey,Ride}.  Furthermore, in the weak probe pulse limit ($a_{0}^2\ll 1$), significant radiation is only generated at the fundamental ($n = 1$). Higher-order odd harmonics start to appear in the spectrum as $a_{0}$ approaches unity.  The highest intensity, however, is still observed at the fundamental. The energy distribution of the backscattered radiation  ($\theta = 0^{\circ}$) at the fundamental harmonic produced from the head-on scattering  ($\phi = 0^{\circ}$) of a single highly relativistic electron  ($\gamma_{0} \gg 1$) with a  probe pulse for which $a_{0}^2\le 1$ is equal to
\begin{equation}
	\frac{\rmd^{2} I_{\rm{single}}}{\rmd\omega_{\rm x} \rmd\Omega} = 16r_{\rm e}m_{ \rm e}c\gamma^2N^2_{0}\frac{\alpha^2_{1}}{a^2_{0}}\left[1-\frac{\alpha_{1}}{2}\right]^2 R(\omega_{\rm x}, \omega_{0})
	\label{backscatter}
\end{equation}
where $\alpha_{1} = a^2_{0}/[4\left(1+a^2_{0}/2\right)]$, $r_{\rm e}= e^2/(4\pi\epsilon_{0}m_{\rm e}c^2)$ is the classical electron radius,  $\epsilon_{0}$ is the vacuum permittivity,  $N_{0}$ is the number of laser oscillations encountered by the electron, and the resonance function $R(\omega_{\rm x}, \omega_{0})$  has the form
\begin{equation}
	R(\omega_{\rm x}, \omega_{0}) = \sinc^{2}{(\omega_{\rm x} - M_{0}\omega_{0})\bar{T}}
	\label{R}
\end{equation}
where $\bar{T} = L_{0}/(2cM_{0})$,  $M_{0} = 4\gamma^2_{0}/(1+a^2_{0}/2)$ is the relativistic Doppler upshift factor, and $L_{0}$ is the spatial laser pulse length \cite{Esarey}.  

If not only a single electron is considered but a distribution which can be described by an energy distribution $f(\gamma)$, the total fundamental radiation spectrum is given by \cite{Schwoerer, Leemans}
\begin{equation}
	\frac{\rmd^{2} I_{\rm T}}{\rmd\omega_{\rm x} \rmd\Omega} = \int  \!  f(\gamma) \frac{\rmd^{2} I_{\rm{single}}}{\rmd\omega_{\rm x} \rmd\Omega}   \rmd \gamma.
\end{equation}
Integrated over the electron energy distribution, the spectrum is then expressed as \cite{Liesfeld, Catravas}
\begin{equation}
	\frac{\rmd^{2} I_{\rm T}}{\rmd\omega_{\rm x} \rmd\Omega} = 8r_{\rm e}m_{\rm e}c\gamma^3N_{0}\frac{\alpha^2_{1}}{a^2_{0}}\left[1-\frac{\alpha_{1}}{2}\right]^2 f(\gamma).
	\label{backscatterTotal}
\end{equation} 
The corresponding photon spectrum, which is calculated from  $\Delta N_{\rm x}/(\Delta\omega_{\rm x} \Delta\Omega) = (N_{\rm b}/\hbar\omega_{\rm x})\rmd^{2} I_{\rm T}/(\rmd\omega_{\rm x} \rmd\Omega)$ 
where $N_{\rm b}$ is the total number of electrons in the beam, is equal to
\begin{equation}
	\frac{\Delta N_{\rm x}}{\Delta \hbar\omega_{\rm x} \Delta\Omega} = \frac{\alpha_{\rm f}}{2\hbar\omega_{0}}N_{0}\gamma \alpha_{1}\left[1-\frac{\alpha_{1}}{2}\right]^2N_{\rm b}f(\gamma).
	\label{NE}
\end{equation}  
where $\alpha_{\rm f}=e^2/(4\pi\epsilon_{0}\hbar c)$ is the fine structure constant \cite{Liesfeld}. The photons are emitted within a narrow cone of half-angle
\begin{equation}
	\Delta\theta_{1} \approx \frac{1}{\gamma_{0}}\sqrt{\frac{1+a^2_{0}/2}{N_{0}}}
	\label{Deltatheta}
\end{equation}
around $\theta = 0^{\circ}$ \cite{Ride}.

Laser electron acceleration using solid targets and linear laser polarization typically produces a broad energy distribution $N_{\rm b}f(\gamma)= \rmd N_{\rm e}/\rmd\gamma$ where $\rmd N_{\rm e}/\rmd\gamma$ is the electron spectrum \cite{Schwoerer}.  This can  be approximately described by an exponentially decaying distribution \cite{Malka,Kruer},
\begin{equation}
	f(\gamma) = f_{0} \exp{\left[-\gamma m_{\rm e}c^{2}/k_{\rm B}T_{\rm e}\right]}
	\label{fgamma}
\end{equation}

\begin{figure}
\begin{center}
	\includegraphics[width=0.7\textwidth]{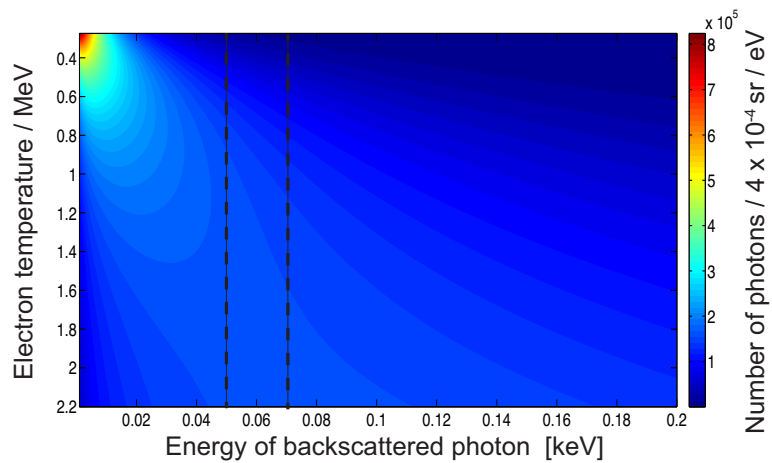}
\end{center}		
	\caption{\label{dNdE_temp}Backscattered photon spectrum as a function of the longitudinal electron temperature $T_{\rm e}$.  The spectrum has been calculated assuming an exponential electron energy distribution with a bunch charge of 30 nC, a laser pulse with a temporal flat-top shape  with a photon energy of 1.55 eV, 13 cycles pulse duration and $a_{0}=1.1$. The black dashed lines mark the spectral range measured in the experiment.}
\end{figure}

\begin{figure}
\begin{center}
	\includegraphics[width=0.7\textwidth]{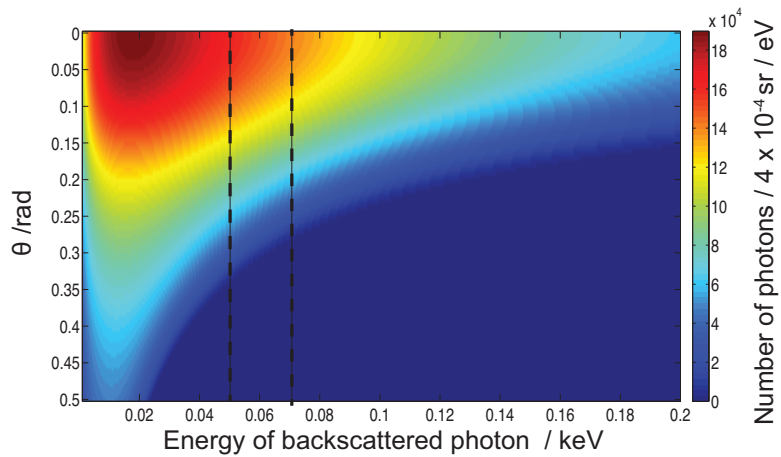}
\end{center}	
	\caption{\label{dNdE_theta}Photon spectrum as a function of the observation angle for a mean electron energy of $k_{\rm B}T_{\rm e} = 1.1$ MeV and the same parameters as in \Fref{dNdE_temp}. The black dashed lines mark the spectral range measured in the experiment}
\end{figure}

Here $\int  \!  f(\gamma) \, \rmd \gamma=1$, $f_{0}$ is a normalization constant \cite{Michel}, $k_{\rm B}$ is the Boltzmann constant and $T_{\rm e}$ the electron quasi-temperature in the propagation direction. For illustration, \eref{NE} is used to compute the backscattered photon spectrum emitted from an electron distribution as described by \eref{fgamma}. \Fref{dNdE_temp} ~shows that at electron temperatures $T_{\rm e}$ of less than about 0.8 MeV, the backscattered spectrum is peaked at low photon energies. At higher electron temperatures, the emission extends toward higher photon energies resulting in a broader and flatter spectrum of the Thomson backscattered radiation.

The spatial distribution of the photon spectrum emitted by an exponential electron distribution with a mean energy of $k_{\rm B}T_{\rm e} = 1.1$ MeV  is shown in \Fref{dNdE_theta}. The emission peak is situated at $\theta = 0$ rad and the central radiation cone has a half-angle of $\Delta\theta_{1} \approx 0.12$ rad as computed from \eref{Deltatheta}. For larger observation angles, the spectrum is restricted to lower photon energies. 

\subsection{The electron sheath}
A relativistic electron population is generated by irradiating the front surface of a thin foil target of thickness $d_{\rm t}$ at normal incidence with an intense pump laser pulse of focal spot size $d_{\rm f}$.  With this configuration, the ponderomotive acceleration mechanism pushes the electron population into and eventually through the target foil. Fast electrons exit at the rear surface leaving a positively charged target behind.  Due to the charge separation, an electric field builds up. As a result, the majority of the electrons of the distribution is forced to return to the target since they are trapped in the induced electric field - forming an electron sheath \cite{Jeackel2}. Assuming a ballistic propagation of these electrons through the target, the divergence half-angle of the beam is found to be $8^{\circ}$ \footnote[1]{The analytical model proposed in \cite{Schreiber} (which is supported by experimental results from \cite{Fuchs,Zepf,Kaluza2})  predicts the dependence of the divergence angle of the electron beam on the energy of the incident laser pulse.  The divergence angle used in this paper is consistent with a 1-J laser like the JETI laser system.} \cite{Honrubia}. From geometrical estimates, the full-width at half maximum (FWHM) radial extent of the electron sheath is computed as \cite{Kaluza2}
\begin{equation}
	w_{\rm ne} = d_{\rm f} + 2d_{\rm t}\tan{8^{\circ}}.	
	\label{wne}
\end{equation}
However, these electrons are pulled back by the arising field and are forced to turn around towards the target \cite{Pukhov}. This phenomenon causes a fountain-like spreading of the electron sheath and hence is called the fountain effect. Once these fountain electrons reach the target front side, they are reflected back to the rear side then they come back again to the front surface and so on \cite{Sentoku}. This recirculation causes the electrons to lose energy and spread out sidewards \cite{Kaluza3}. Consequently, a much larger radial extent of the electron sheath than predicted by \eref{wne} is expected \cite{Jeackel2}. The electron sheath then consists of a very dense center while the density rapidly drops in the outer regions \cite{Kaluza3}.   

For the interaction of this electron sheath with a counterpropagating probe pulse, it can be assumed that the scattering process is dominated by the dense center of the electron sheath. And since the electron temperature is also maximum at the sheath center \cite{Kaluza3}, the contribution of  fountain electrons to the detected spectrum is minimal. In addition, the spatial distribution of the backscattered  radiation is centered in the electron's forward direction i.e. along z' making an angle $\phi$ with respect to the z-axis. For the fundamental harmonic, the emission is confined to a cone of half-angle $\Delta\theta_{1}$. If the detection system (solid angle of $\Delta\Omega = 4 \times 10^{-4}$sr with subtended half-angle of $\theta_{\rm det} = 0.01$ rad) is situated along the z-axis centered at $\phi = 0$ rad, the backscattered radiation from electrons with interaction angle $\phi \geq \Delta\theta_{1} +\theta_{\rm det} $ is no longer measured. This further minimizes the impact of fountain electrons to the Thomson backscattered signal. Therefore,  in this paper it is sufficient to consider only the backscattered radiation from a head-on scattering geometry. 

\section{Experimental Setup}
\label{sec:ExperimentalSetup}
\subsection{Layout}

\begin{figure}
\begin{center}
	\includegraphics[width=0.65\textwidth]{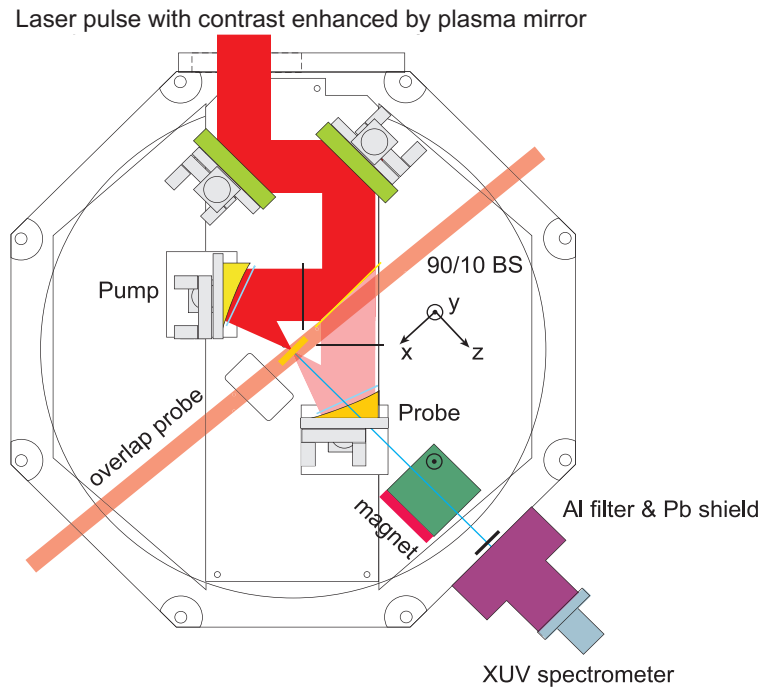}
\end{center}		
	\caption{\label{setup}Experimental setup. The main JETI pulse was split into two by a 3-mm thick dielectric-coated 90/10 beam splitter. Both pulses were then focused at normal incidence onto the aluminium foil target  using two \textit{f}/2 $45^{\circ}$ off-axis parabolic (OAP) mirrors.  The backscattered radiation was detected using a XUV spectrometer and an Andor$^{\rm{TM}}$ XUV CCD camera arrangement. Spatio-temporal overlap between the two pulses was established using the shadowgram of the interaction region and by the 0$^{th}$-order image on the CCD camera.} 
\end{figure}

The experiment was carried out at the Jena Titanium:Sapphire TW laser system (JETI). It provided linearly polarized pulses at a central wavelength of $\lambda_{\mu} = 0.8$ $\mu$m ($\hbar\omega_{0} = 1.55 $ eV) with $\tau = 30$ fs duration.  A single pass plasma mirror was employed to improve the contrast ratio between the main pulse and the pre-pulses (occurring 150 ps before the main pulse) from $10^{-6}$ to $10^{-9}$ and the ASE background from $10^{-9}$ to $10^{-12}$ \cite{Roedel2}.  A diagram of the experimental setup is presented in \Fref{setup}.

The main pulse of the JETI laser, with beam diameter of $d_{\rm beam} = 6.0$ cm, was divided into a pump and a probe pulse by a 3-mm thick dielectric-coated 90/10 beam splitter. 90\% of the pulse energy was utilized as the pump pulse while the remaining 10\% was employed as the probe pulse. Each of these pulses was focused at normal incidence onto the target by two $45^{\circ}$ off-axis gold-coated parabolic mirrors of effective focal length \textit{f}$ = 12.0$ cm.  Using this configuration, the pump pulse was focused to a focal spot of A$_{\rm{Pump}} = 8.9~\mu$m$^2$ (FWHM) to a peak intensity of $I_{\rm{Pump}} = 1.9\times 10^{19}$ W/cm$^2$ corresponding to a normalized vector potential of $a^{\rm{Pump}}_{0} = 3.0$.  On the other hand, the probe pulse was focused to a spot of A$_{\rm{Probe}} = 7.5~\mu$m$^2$ (FWHM)\footnote[2]{The spot size of the pump pulse is slightly larger than the probe pulse due to its reflection from the beam splitter. The probe pulse, in contrast, is transmitted. Therefore, imperfections on  the beam splitter surface do not significantly alter the beam quality.} to an intensity of $I_{\rm{Probe}} = 2.8\times 10^{18} $ W/cm$^2$ with $a^{\rm{Probe}}_{0} = 1.1$.

The intense pump pulse accelerates electrons into the $d_{\rm t} = 2.4~\mu$m thick aluminium foil and the photons from the probe pulse impinging on the target rear surface are backscattered off the electrons exiting the rear surface.  The backscattered radiation was collected through a 3-mm hole  in the off-axis parabolic mirror of the probe.  A 0.5-T magnet located on axis at the back of the parabolic mirror was then employed to prevent accelerated charged particles from propagating through the detection system.  An XUV spectrometer employing a toroidal mirror with a collection solid angle of $\Delta\Omega = 4\times10^{-4}$sr and a binary transmission grating made of free-standing gold bars with 2000 lines/mm  together with an Andor$^{\rm{TM}}$ XUV CCD camera make up the detection system in this experiment. Background low-energy plasma emission as well as ambient radiation were blocked using two 200 nm thick aluminium filters placed in front of the spectrometer. Additional lead shielding was also placed at the entrance of the spectrometer to enhance the suppression of background radiation. This gives an effective detection range of 52 eV to 72.7 eV. The setup, excluding spectrometer and camera, was placed in an octagonal vacuum chamber where a pressure of about $10^{-5}$ mbar was applied. 

\subsection{Spatio-temporal Overlap}

\subsubsection{Alignment in the direction of the \textit{y}, \textit{z} and \textit{t} axes}
An additional temporally synchronized overlap probe was utilized to back-light the pump-probe interaction region [see  \Fref{setup}]. It created a shadowgram of the ionization front produced by each pulse individually when fired through air.  The shadowgram was imaged onto a CCD camera.  The overlap in the \textit{y}- and  \textit{z}- axes of the laser foci was obtained by intersecting the region with the peak plasma self-emission of each ionization front. To establish the spatial overlap, the probe OAP focus was translated towards the pump OAP focus. Then the corresponding temporal overlap was adjusted by compensating the delay introduced during the probe OAP alignment with an appropriate translation of the beam splitter. This was done by noting that shifting the beam splitter by a distance of $\Delta l$ normal to its surface makes the pump pulse cover an additional optical path of  $\Delta\sigma =\sqrt{2} \Delta l$. Furthermore, the delay stage of the overlap probe was scanned to mark the onset of each ionization front. Then the position of the beam splitter was again adjusted to balance the difference in the arrival time of both pulses.  This was then used to set the temporal delay of $\Delta t = 0$ between the arrival of the two pulses at the target position with a resolution of approximately 60 fs.  

\subsubsection{Alignment in the direction of the \textit{x} and \textit{y} axes}
The alignment of the laser foci in direction of the \textit{x}- and \textit{y}- axes was established by noting separately the pixel positions of the $0^{th}$ order diffraction of the emitted light from each OAP as measured by the XUV CCD. The optimum overlap of the foci with respect to these axes was then achieved by matching these pixel positions. With this approach, the foci can be overlapped with an accuracy of about $13~\mu$m in both transverse directions which corresponds to half the pixel size.

\subsubsection{Size of the interaction region}
The radial extent of the electron sheath created by the pump laser pulse is determined from \eref{wne} and is found to be $w_{\rm ne} = 4~\mu$m. This corresponds to a FWHM electron sheath area of $A_{\rm{Sheath}} = 13~\mu$m$^2$. This matches well with the probe pulse focal spot A$_{\rm{Probe}} = 7.5~\mu$m$^2$ and is already sufficient for observing the backscattering process.   Moreover, due to the fountain effect an even larger electron sheath spread is expected. In fact, J\"ackel et al. observed that the experimentally deduced radial extent is about a factor of 2.5 times larger than the estimated value \cite{Jeackel2}. 

\section{Experimental Results}
\label{sec:Results}

\begin{figure}
\begin{center}
	\includegraphics[width=0.65\textwidth]{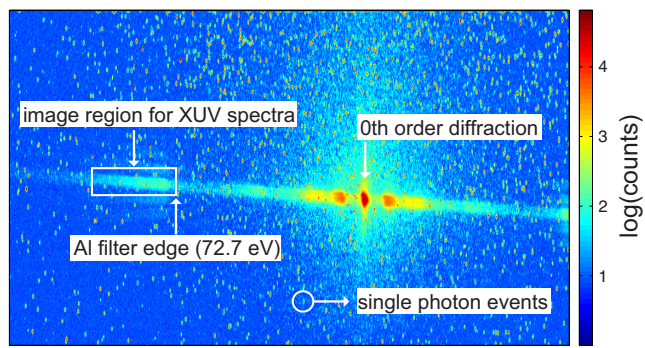}
\end{center} 	
	\caption{\label{raw}Raw CCD image of the XUV spectrometer in false-colours taken during a single laser shot. A logarithmic colour table was used for better image rendering. To increase the spectral range accommodated by the spectrometer, the grating was slightly rotated resulting in a minor tilt of 	the XUV spectral trace. The aluminium filter edge at 72.7 eV in combination with the $0^{th}$ 	order diffraction were used to establish the pixel to wavelength calibration.}
\end{figure}
  
\begin{figure}
\begin{center}
	\includegraphics[width=0.65\textwidth]{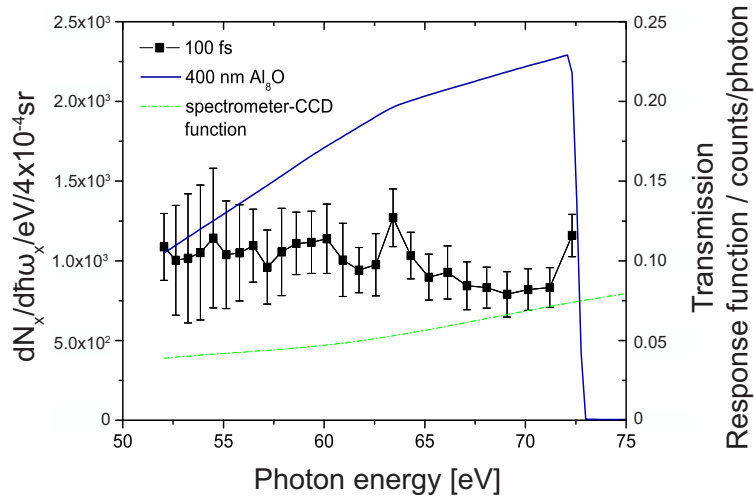}
\end{center}	
	\caption{\label{Al2m100fs}XUV photon spectrum measured for a positive pump-probe delay of 100 fs. The error bars 	indicate the standard error using 16 subsequent laser shots. The filter transmission (blue line) and the response function (green dashed line) are included in the spectrum.}
\end{figure}

A typical raw image produced by the Thomson- scattered radiation from a single laser shot is shown in   \Fref{raw}. The raw XUV spectrum, represented by the slightly tilted and spectrally dispersed signal in the middle of the image, is overlaid with single and multiple pixel events. A numerical filter was applied to the image to remove isolated pixel events, thus separating the raw XUV spectrum. Using the filter transmission and the response function of the spectrometer-CCD system the number of photons is deduced from the CCD signal. Given in \Fref{Al2m100fs} is the XUV spectrum generated when the pump pulse arrived on the target position 100 fs prior to the probe pulse. 

\begin{figure}
\begin{center}
	\includegraphics[width=0.65\textwidth]{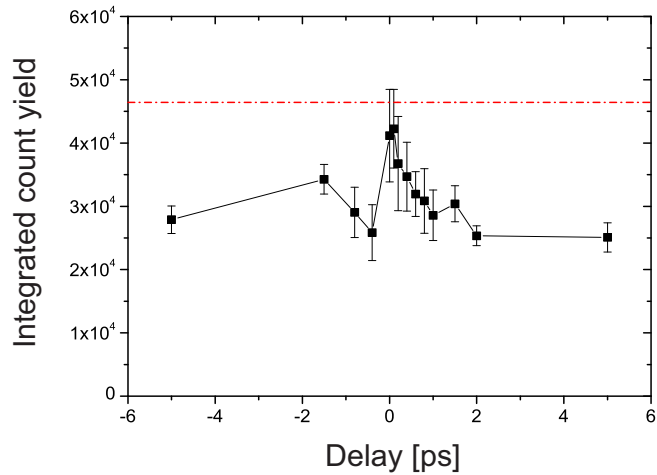}
\end{center}	
	\caption{\label{Al2mcountyield}Dependence of the integrated count yield  of the XUV signal (energy range between 52 eV and 72.7 eV) on the pump-probe delay. The error bars indicate the standard error of the integrated count yield for 16 laser shots. The red dashed line indicates the sum of the XUV count yield due to the pump and the probe pulse.}
\end{figure}

A direct evidence that Thomson backscattering is the dominant process in the generation of high energy photons is the dependence of the XUV signal on the pump-probe delay.
Shown in \Fref{Al2mcountyield} is the summation of the number of counts of the raw XUV spectra over the energy range between 52 eV and 72.7 eV for different pump-probe delay.  A negative delay indicates that the probe pulse hits the target before the pump pulse while the opposite is true for a positive delay. \Fref{Al2mcountyield} highlights the occurrence of a peak at 100 fs and the gradual decrease in the count yield with increasing positive delay. At the 100-fs delay setting the detected XUV signal is enhanced by Thomson-backscattered photons. This can be explained by the simultaneous presence of the highest electron density and temperature at this instance in the interaction area.  At later times, the rear-surface electron sheath continually expands longitudinally and in the radial direction \cite{Jeackel2}. This transverse plasma expansion and the associated adiabatic cooling of the electron population causes the steady decrease in the signal for increasing positive delays.  It is worth emphasizing that since the positive pump-probe delays have femtosecond time-scales, the strong dependence of the XUV yield on the delay can only be due to the Thomson backscattering process.  Surprisingly, an appreciable signal yield is also observed for negative pump-probe delays.  In this case the intense probe pulse generates electrons itself that are accelerated in the opposite direction as compared to those originating from the interaction of the target with the main pulse (i.e. they are moving from the target rear side to the target front side). These electrons then form a preplasma on the target front side into which the more intense pump pulse is focused. At a delay of -400 fs,  the preplasma created by the probe beam is still hot. However, for a delay of -1.5 ps, the preplasma electrons will have cooled down allowing an optimum energy coupling with the pump laser pulse. Therefore the more intense pump pulse accelerates the electrons towards the rear side of the target, generating a considerable signal from broadened line emissions and Bremsstrahlung radiation \cite{Rousse,Chichkov}. Indeed, \Fref{Al2mcountyield} shows an elevated count yield at this delay.

Although the results above are consistent with Thomson backscattering, the distribution of the backscattered signal can only be unambiguously determined by isolating the background signal.  In this particular case, however, it was not straightforward to discriminate the exact background signal. This comes from the fact that when both pulses interact with the target they essentially share a common volume of the aluminium foil where the background and backscattered signal are produced from. Whereas, when only either of the pulses interacts with the foil target, the entire bunch of accelerated electrons generates the background signal. It thus follows that the combined signal yield of the pump and the probe is expected to be larger than the total backscattered signal yield.  The red dashed line in \Fref{Al2mcountyield} indicates the sum of the signal yield of the pump and probe pulses. As predicted, it is found to be higher than the signal obtained when both pulses interact with the target.  

\begin{figure}
\begin{center}
	\includegraphics[width=0.65\textwidth]{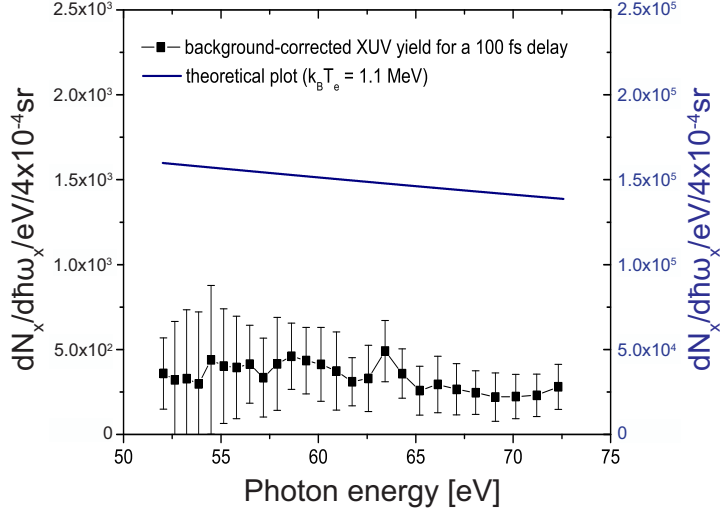}
\end{center} 	
	\caption{\label{Al2m100fs_backsub} The background-corrected XUV spectrum for a pump-probe delay of 100 fs showing statistically significant Thomson backscattered signal for photon energies greater than 52 eV (black square).  The background signal is estimated from the average of the spectrum from -5 ps and +5 ps delays. The error bars indicate the standard error from 16 laser shots.  A theoretical plot of the XUV spectrum obtained for an electron distribution with a mean kinetic energy  of $k_{\rm B}T_{\rm e} = 1.1$ MeV (blue line) exhibits the same shape as the background-corrected spectrum. The difference in the absolute photon yield between the two plots can be explained by the imperfect overlap of the electron beam with the laser pulse in the experiment.}
\end{figure}	

An estimate of the background radiation can be found by using the average of the signal from the -5-ps and +5-ps delays. The resulting XUV spectrum for 100 fs delay after background subtraction and a theoretical plot of the XUV spectrum at an electron kinetic energy of 1.1 MeV are shown in \Fref{Al2m100fs_backsub}. The backscattered spectrum is broad which obviously arises from the broad electron energy distribution as noted in \Fref{dNdE_temp}.  This is expected due to the ponderomotive electron acceleration mechanism which is likely to be dominant under our experimental conditions. Note that at normal incidence of the pump pulse on the target, effects like resonance absorption or Brunel heating are strongly suppressed \cite{Brunel}. It is also evident that Thomson-backscattered photons are produced. However, the signal has statistical significance only at energies greater than 52 eV. This is consistent with backscattering from the escaping hot electron population:  The ponderomotive scaling law predicts that electrons oscillating in the field of an incident laser will gain a mean kinetic energy of $k_{\rm B}T_{\rm e} = (\gamma_{0}-1)m_{\rm e}c^2$ where $\gamma_{0}= \left(1+I_{\rm{Pump}}\lambda^2_{\mu}/1.37 \times 10^{18} \textrm{W} \mu\textrm{m}^2/\textrm{cm}^2\right)^{1/2}$ \cite{Wilks}. Considering our pump pulse intensity, the mean kinetic energy of the electrons correspond to $k_{\rm B}T_{\rm e} = 1.1$ MeV  with $\gamma_{0}=3.0$.  A small fraction of these hot electrons escape from the target rear surface charging up the target to a potential of the order of  $k_{\rm B}T_{\rm e}$ \cite{Kar}. The number of escaped electrons $N_{\rm esc}(t)$ and the self-capacitance of the target $C_{\rm T}(t)$ then dictate the temporal evolution of the target potential $V(t)$ such that $V(t) = N_{\rm esc}(t)e/C_{\rm T}(t)$ where $N_{\rm esc}(t) = N_{\rm b}\exp(-eV(t)/k_{\rm B}T_{\rm e})$ and $C_{\rm T}(t) \approx 8\epsilon_{0}(r_{\rm f}+0.75ct)$ assuming that the charging expands from the laser spot size of radius $r_{\rm f}$ with a radial velocity of the charge wave approximated at $0.75c$ \cite{Kar, Myatt, McKenna}.  Consequently, once a sufficient number of electrons have escaped, the bulk of the hot electron population becomes electrostatically trapped on the target rear surface and only those that are   energetic enough to overcome the developing potential barrier can escape \cite{Borghesi}. Therefore, backscattering from the escaping hot electron component having $\gamma_{0} \ge 3.0$ will produce photons at the Doppler upshifted energies of $\hbar\omega_{\rm x} \ge $ 56 eV. Furthermore, since the escaping hot electrons scatter more strongly than their lower energy counterparts, they contribute dominantly to the backscattered spectrum.  The consistency between the experimentally observed behaviour of the XUV spectrum and theoretical predictions indicate the presence of Thomson backscattering in the photon production.

\section{Conclusion}
\label{sec:Conclusion}
An all-optical, laser-based setup was presented for investigating Thomson backscattering. At the present conditions, the photon spectrum produced is broad due to the broad electron distribution.  A statistically significant photon yield is detected at photon energies greater than 52 eV which corresponds to backscattering from the escaping hot electron population.  In addition, the XUV signal exhibits a strong dependence on the difference between the arrival time of the pump and the probe pulses at the target position. A maximum XUV signal is obtained when the pump pulse interacts with the target 100 fs before the probe pulse. The XUV signal also decreases with increasing positive pump-probe delays. The behaviour of the XUV spectra, particularly at increasing positive delays clearly indicate that the source of these high-energy photons is Thomson backscattering. It was concluded that the highest flux and energy in the Thomson backscattered photon signal is generated by the rear-surface electron sheath having both the highest density and also temperature. Since these parameters are also of great importance for the acceleration of ions from the rear surface of thin foils \cite{Honrubia}, the spectral and temporal behaviour of the Thomson backscattered signal has the potential to provide additional information about the temporal evolution of the ion-acceleration fields. 

Optimization of the experiment will involve the use of electron sheets with a quasi-monoenergetic distribution. Using ultra-thin diamond-like carbon foils being accelerated in the radiation pressure acceleration regime might be a future alternative to produce both monoenergetic electron and - as a consequence - XUV and even x-ray spectra.    

\ack{This work was supported by the Deutsche Forschungsgemeinschaft (SFB/TR 18) and the German Federal Ministry of Education and Research (BMBF) under contract 05K10SJ2. C.R. acknowledges the support from the Carl Zeiss Foundation. We acknowledge the contributions of Burgard Beleites and Falk Ronneberger to this work by maintaining the JETI laser system.}

\end{document}